\documentclass[11pt]{article}

\usepackage[final]{acl}

\usepackage{times}
\usepackage{latexsym}
\usepackage{amsmath}
\usepackage{booktabs}
\usepackage{array}

\usepackage[T1]{fontenc}

\usepackage[utf8]{inputenc}

\usepackage{microtype}

\usepackage{inconsolata}
\usepackage{pifont}

\usepackage{graphicx}
\usepackage[most]{tcolorbox}
\usepackage{listings}
\tcbuselibrary{listings,breakable}

\newtcblisting{taskbox}[1][]{
    enhanced,
    listing only,
    colback=gray!3,
    colframe=black!70,
    colbacktitle=black!75,
    coltitle=white,
    fonttitle=\small\bfseries,
    boxrule=0.8pt,
    arc=1.5mm,
    outer arc=1.5mm,
    left=2mm,
    right=2mm,
    top=1.5mm,
    bottom=1.5mm,
    boxsep=1mm,
    before skip=6pt,
    after skip=8pt,
    listing options={
        basicstyle=\ttfamily\footnotesize,
        columns=fullflexible,
        keepspaces=true,
        showstringspaces=false,
        breaklines=true
    },
    #1
}

\newcommand{\ie}{\emph{i.e.}}
\newcommand{\eg}{\emph{e.g.}}

\title{Lab2Sim: Turning Real-World Experiment Videos into \\ Interactive Virtual Experiments}

\author{
Xusen Hei$^{1,*}$, Jiali Chen$^{1,2,*}$, Jinyu Yang$^{1}$, Yi Cai$^{1,\dagger}$ \\
$^{1}$South China University of Technology \\
$^{2}$The Hong Kong Polytechnic University \\
{\tt\small sexusenhei@mail.scut.edu.cn}
}

\begin{document}
\maketitle
\begin{abstract}

Science experiment videos demonstrate how experiments are performed, but offer no direct interaction for learners. Turning them into virtual experiments requires not only reproducing the observed apparatus and state changes, but also enforcing the preconditions governing each operation. We introduce Lab2Sim, a benchmark of 218 science experiment videos with evidence-grounded specifications and 11,881 rubric items. To address this task, we propose EduVideo2Code, which extracts experimental procedures hierarchically, encodes them as explicit state machines to guide code generation, and refines the resulting pages through execution-based verification. Experiments show that existing multimodal models readily generate runnable pages but struggle to faithfully reconstruct experimental apparatus. EduVideo2Code consistently improves all four evaluation dimensions across two backbone models, with apparatus fidelity gains of 24.15 points\footnote{Our dataset and code will be released at the project page: \url{https://black1948.github.io/lab2sim}}. 
\end{abstract}

\section{Introduction}
\begin{figure*}[t]
\centering
\includegraphics[width=\textwidth]{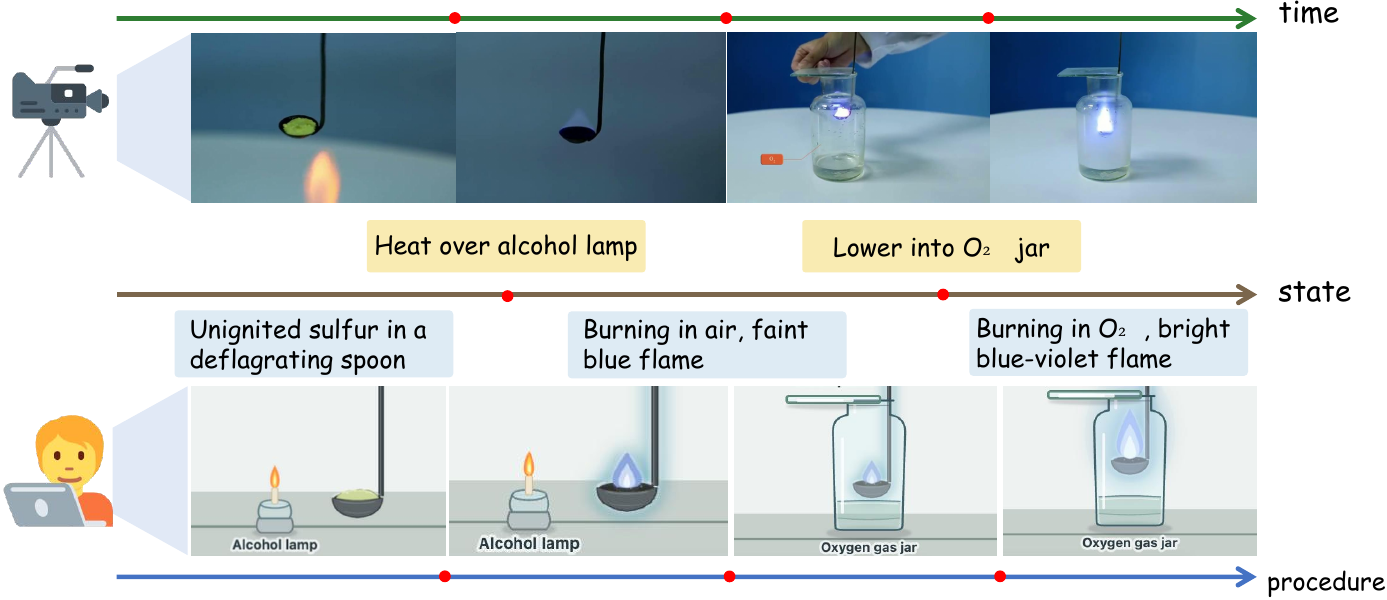}
\caption{Motivating example of experiment video-to-simulation generation.}
\label{fig:intro}
\end{figure*}

Scientific experiment videos provide an accessible way to learn how experiments are performed, but learners can only observe rather than conduct them. Automatically converting these videos into browser-based virtual experiments could bridge this gap and make hands-on scientific learning more accessible. However, such conversion requires two capabilities: fidelity and operability. Fidelity requires reproducing the apparatus, experimental procedure, and observable state transitions shown in the video, while operability requires enabling learners to perform each step under the correct preconditions and observe the expected results.

Existing work addresses these requirements only partially. Front-end code generation methods reconstruct webpages from visual references, including screenshots, design specifications, and screen recordings, with recent benchmarks evaluating their functional correctness and interactive behavior \cite{design2code,web2code,sketch2code,webgen-bench,webgen-agent,frontendbench,webgrader,video2code,iwr-bench,animation2code,webvr}. However, reconstructing interface transitions does not require inferring the physical states of experimental materials or the procedural constraints governing laboratory operations. Scientific video understanding methods recognize experimental events and reason about scientific phenomena \cite{video-mmlu,mmvu,video-mmmu,expvid,physlab,scivid,expstar}, but primarily produce textual answers or explanations rather than interactive environments. Thus, a key challenge remains: \textit{how can a system translate what happens in an experiment video into a virtual experiment that is both faithful to the demonstration and operable by learners?}

We address this question with Lab2Sim, a benchmark of 218 chemistry, physics, and biology experiment videos, each paired with a video-grounded specification of its apparatus, ordered steps, preconditions, and state transitions. From these specifications, we derive 11,881 rubric items to evaluate both fidelity and operability by executing the generated pages in a browser. Building on this benchmark, we propose EduVideo2Code, a compile--verify--repair framework that bridges video understanding and interactive implementation. The framework first extracts experimental procedures through coarse-to-fine video analysis, then compiles them into an explicit state machine that guides code generation, and finally verifies and repairs the generated page through browser execution. By making experimental logic explicit and grounding refinement in observed behavior, EduVideo2Code aims to move beyond generating pages that merely run toward virtual experiments that faithfully reproduce real procedures.

Experiments with six multimodal models reveal a substantial gap between executability and fidelity: although more than 93\% of directly generated pages are runnable, no model achieves 60\% apparatus fidelity. EduVideo2Code improves all four evaluation dimensions on both tested backbone models, with gains of up to 24.15 points in apparatus fidelity. Ablations further show that all three stages contribute to performance. These findings demonstrate that converting experiment videos into usable learning environments requires not only code generation, but also explicit procedural modeling and execution-based verification.

Our contributions are as follows:
\textbf{(i)} We introduce Lab2Sim, a benchmark of 218 real-world experiment videos with evidence-grounded specifications and 11,881 executable rubric items for evaluating the fidelity and operability of generated virtual experiments. 
\textbf{(ii)} We propose EduVideo2Code, a framework that combines hierarchical procedure extraction, state-machine-driven code generation, and execution-based verification and repair.
\textbf{(iii)} We establish the gap between page executability and experimental fidelity through evaluations of six multimodal models, and demonstrate consistent improvements from EduVideo2Code across four evaluation dimensions.

\section{Related Work}
\subsection{Front-End Code Generation}
Front-end code generation has moved from reproducing static appearance to generating interactive interfaces. Early work reproduces webpage layouts from screenshots, sketches, or design specifications \cite{design2code,web2code,sketch2code}. Later benchmarks evaluate functional correctness by executing the generated pages and interacting with them \cite{webgen-bench,webgen-agent,frontendbench,webgrader}. More recent work takes interface recordings as input and recovers the states, actions, and transitions they show \cite{video2code,iwr-bench,animation2code,webvr}.

Related works synthesizes programs or reconstructs environments from physical demonstrations \cite{demo2code,poe-world,ego2world,foldingagent,visphyworld}. In these settings, the target behavior is either displayed directly by an interface or is not governed by a scientific procedure. EduVideo2Code instead targets science experiments. It infers an experiment state machine from a physical demonstration and checks the generated page against the states observed in the video.

\subsection{Scientific Experiment Understanding}
Scientific video benchmarks study reasoning and question answering over educational videos \cite{video-mmlu,mmvu,video-mmmu}. Recent datasets focus on fine-grained perception and reasoning in real-world experiments \cite{expvid,physlab,scivid}. Related work on instructional videos studies step localization and temporal action understanding \cite{youcook2,howto100m}, and generative methods produce textual explanations of experiments \cite{expstar}. These methods output labels, answers, or text. None of them produces an environment in which the experiment can be performed. EduVideo2Code fills this gap by converting experiment videos into interactive virtual experiments whose states change in response to the learner's operations.

\section{Lab2Sim}
\label{sec:benchmark}

\subsection{Task Definition}
\label{sec:benchmark-task}

\begin{figure}[t]
\centering
\includegraphics[width=\columnwidth]{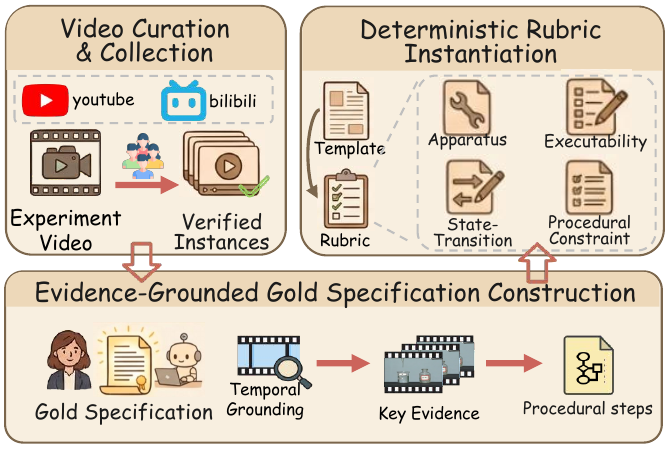}
\caption{Overview of Lab2Sim construction. Videos are collected and screened, a gold specification grounded in video frames is built for each video, and fixed templates turn the specification into rubric items.}
\label{fig:bench-overview}
\end{figure}

Lab2Sim evaluates whether a system can convert a real-world science experiment video into a virtual experiment that runs in a browser. Each instance provides a source video $V_i$ and a generation instruction as input.

The system outputs a webpage $W_i$, which implements the virtual experiment; we refer to it as the \textit{page}. The page should contain the apparatus shown in the video, support the demonstrated steps, enforce their preconditions, and display the resulting state transitions. For evaluation, each instance also has a gold specification $G_i$, built independently of the evaluated systems, and a set of rubric items $R_i$ derived from $G_i$. Neither is shown to the systems. We score $W_i$ by checking whether it satisfies each item in $R_i$. The evaluation targets the experiment's content and behavior rather than pixel-level appearance, and it covers the four dimensions described in Section~\ref{sec:benchmark-dimensions}.

\subsection{Video Collection}
\label{sec:benchmark-collection}

We collected science experiment videos in chemistry, physics, and biology from public instructional and demonstration content on Bilibili and YouTube. From more than 500 candidates, we kept 218 after manual screening. We kept a video only if its procedure, apparatus, and observable state transitions could be reliably identified from the footage. We excluded duplicate recordings of the same experiment, videos below 720p, experiments that were overly complex or heavily occluded, and videos that only gave verbal explanations or showed final results without the procedure. Each kept video corresponds to one virtual experiment.

\subsection{Evidence-Grounded Annotation}
\label{sec:benchmark-annotation}

For each experiment, we build a structured gold specification. It records the apparatus, the initial conditions, the ordered steps, the preconditions of each step, and the observable state transitions. We use \textit{apparatus} for both equipment and materials, \textit{step} for a manipulation or observation demonstrated in the video, \textit{operation} for the page's implementation of a step, and \textit{state transition} for the change in an object's state that a step causes, such as a solution turning from colorless to pink. An assisting model from the GLM-5.3 series drafts each specification, and human reviewers then verify every draft against the source video, which takes about 13 minutes per video on average. 

The draft is produced from coarse to fine. A first pass over the whole video establishes the experimental context and lists candidate steps and their state transitions. A second pass inspects each relevant interval to fix event boundaries and describe the state transitions, and it samples additional frames whenever the evidence for a key event is unclear. Appendix~\ref{app:annotation-guidelines} gives the full guidelines.

Every annotated element is linked to one or more source frames or to a time interval. Each step records its target object, its step type from a fixed vocabulary (e.g., pour or heat), its time span, and its pre- and post-states when they can be identified. Each precondition records the state that the step requires (e.g., \texttt{stopcock = open}), together with a script test that detects when the step is attempted without it. A precondition is labeled \texttt{video\_visible} when the required state can be seen in the video before the step, for example the demonstrator opening the stopcock before the titrant flows. The footage shows only that the precondition held. The requirement that the page block the step when the precondition does not hold comes from operability, not from the video. Each state transition is recorded as an update to one attribute of a named object, for example \texttt{flask.solution.color}. Each element also carries one of three evidence labels: \texttt{video\_visible} for facts that can be seen directly in the video, \texttt{apparatus\_inferred} for properties inferred from the visible apparatus and context, and \texttt{domain} for textbook knowledge that the footage does not show. Claims with no visual evidence, no justified inference, and no textbook basis are discarded. Domain claims are kept, but their label ensures that they are never counted as facts from the video.

Fixed templates convert the reviewed gold specification into rubric items. Apparatus entities, steps, preconditions, and state transitions (including initial conditions) yield identity, execution, constraint, and state items, respectively (Table~\ref{tab:instantiation-templates} in Appendix~\ref{app:instantiation-templates}). In addition, we generate \textit{violation probes}, which skip a required earlier step, swap the order of two steps, use an excessive dose, or apply a step to the wrong object. Unlike constraint items, probes are not tied to an annotated precondition; deciding that the violation is invalid requires domain knowledge, so probes are labeled \texttt{domain}. Each rubric item keeps a pointer to its source field and its video evidence, so every item can be traced back to the footage. Human reviewers check the gold specification for completeness and evidential support, and they check each rubric item for validity. The rubric is finalized only after both checks (Appendix~\ref{app:quality-control}).

\subsection{Rubric Evaluation Dimensions}
\label{sec:benchmark-dimensions}

The rubric covers four dimensions. \textit{Apparatus Fidelity (AF)} checks whether the apparatus of the experiment is present and correctly identified; it does not score spatial layout. \textit{Procedure Execution (PE)} checks whether each annotated step can be performed on the correct object and completes as specified. \textit{Procedural Constraint Compliance (PCC)} checks whether the page enforces each precondition, that is, whether it blocks a step whose precondition does not hold and shows feedback. \textit{State-Transition Fidelity (STF)} checks whether the initial states, and the states after each step, match those in the video. AF and STF thus measure fidelity, while PE and PCC measure operability. Identity, execution, constraint, and state items count toward AF, PE, PCC, and STF, respectively.

Each scored rubric item belongs to exactly one dimension, determined by its source field, and violation probes count toward PCC. The items describe the experiment's content and behavior rather than its implementation, so functionally equivalent pages should receive the same score. Execution and state items are checked by replaying the annotated steps. Constraint items are checked by attempting a step before its precondition holds and checking that the page blocks it and shows feedback.

A page must run in the browser before its rubric items are scored, and whether it runs is reported separately as the Runnable Rate. The main scores include items based on \texttt{video\_visible} elements and the violation probes. The other items based on apparatus inference or domain knowledge are kept in the released rubric but excluded from the main scores.
Section~\ref{sec:experiments} describes the execution protocol and scoring.

\begin{table}[t]
\centering
\caption{Statistics of Lab2Sim.}
\label{tab:benchmark-statistics}
\small
\begin{tabular}{lr}
\toprule
Statistic & Value \\
\midrule
Instances (videos) & 218 \\
Discipline (chemistry / physics / biology) & 89 / 87 / 42 \\
Mean video duration & 135.8 s \\
Mean steps per instance & 7.1 \\
Mean state transitions per instance & 23.2 \\
Mean rubric items per instance & 54.5 \\
Total rubric items & 11,881 \\
\bottomrule
\end{tabular}
\end{table}

\subsection{Benchmark Statistics}
\label{sec:benchmark-statistics}

Table~\ref{tab:benchmark-statistics} summarizes Lab2Sim. It contains 218 instances across chemistry, physics, and biology. On average, each instance has 7.1 steps and 23.2 annotated state transitions. In total, Lab2Sim contains 11,881 rubric items, and 54.5 per instance.

\section{EduVideo2Code}
\subsection{Overview}
\begin{figure*}[t]
\centering
\includegraphics[width=\textwidth]{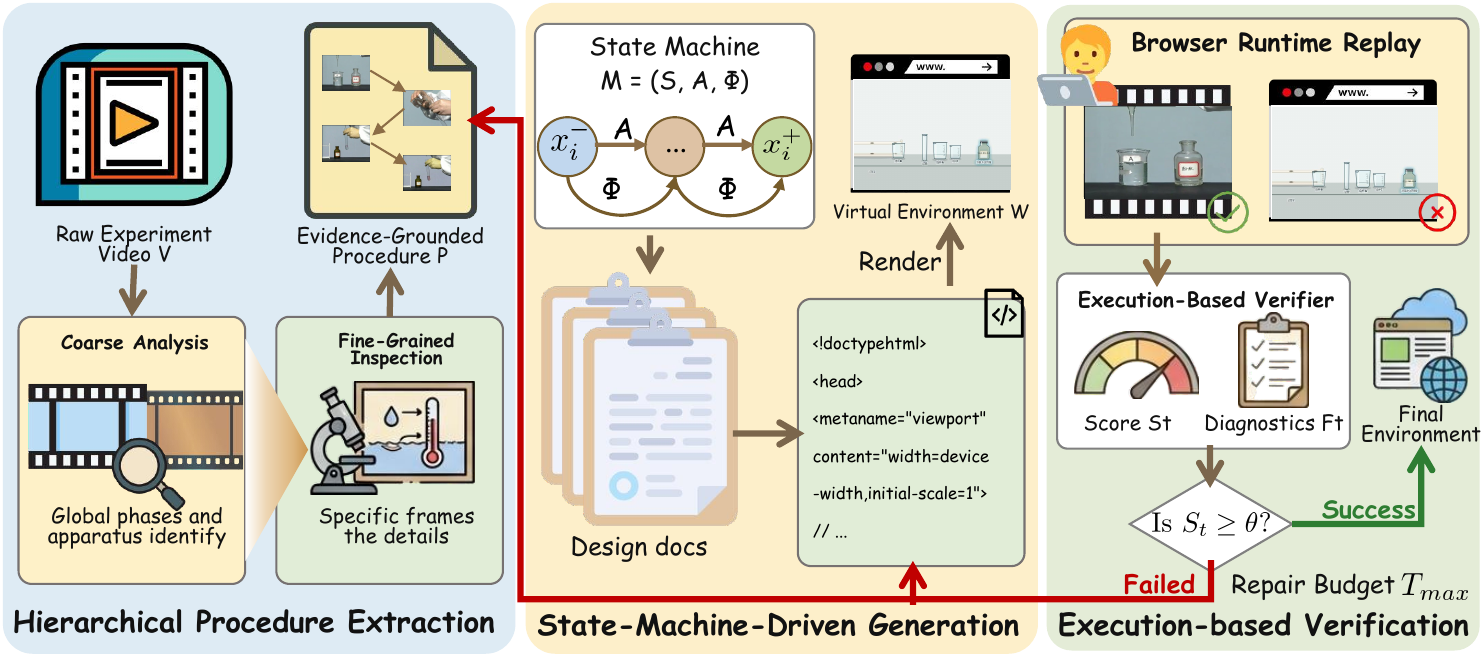}
\caption{Overview of EduVideo2Code. HPE extracts an evidence-grounded procedure $P$ from the video; SMG compiles $P$ into a state machine $M$ and generates the page; EBV runs the page in a browser and, if its score is below the threshold, triggers targeted repair until the page is accepted or the repair budget is used up.}
\label{fig:method-overview}
\end{figure*}

EduVideo2Code converts a real experiment video into an interactive virtual experiment through a compile--verify--repair loop (Figure~\ref{fig:method-overview}). We drop the instance index $i$ in this section. Hierarchical procedure extraction (HPE) recovers the experimental steps, their state transitions, and the supporting video evidence. State-machine-driven generation (SMG) compiles this procedure into an experiment state machine and uses it to generate the page. Execution-based verification (EBV) runs the page in a browser, replays the procedure, tests the preconditions, and compares the resulting states with the video. Any detected error triggers a targeted repair of the procedure, the state machine, or the apparatus, and verification repeats.

\subsection{Hierarchical Procedure Extraction}
Given an experiment video $V$, HPE extracts a procedure $P$, which lists the experimental steps and their state transitions and links each of them to supporting frames. A single pass over the video has two problems: uniform frame sampling can miss brief state transitions, and the model may add steps that the video does not show. We therefore work from coarse to fine. A first pass over the full video identifies the main experimental phases, the apparatus, and the order of steps, and it marks ambiguous segments. We then segment the video at the phase boundaries found in the first pass and inspect the marked segments with denser sampling and high-resolution frames. This second pass targets fine details such as instrument readings, liquid levels, and connections between pieces of apparatus. Because every step in the output must cite supporting frames, steps that the video does not show are less likely to enter the procedure.

The resulting $P$ is the input to SMG and the reference for EBV.

\subsection{State-Machine-Driven Generation}
Generating the page directly from $P$ would mix the experimental logic with scene layout and interaction code. SMG instead first compiles $P$ into an experiment state machine $M=(S,A,\Phi)$. Each state in $S$ assigns a value to every tracked object attribute, such as \texttt{flask.solution.color}. $A$ is the set of operations, each implementing one step of $P$, and $\Phi$ gives the precondition of each operation. When the precondition of an operation $a_k \in A$ holds, the operation moves the experiment from its pre-state $x_k^- \in S$ to its post-state $x_k^+ \in S$. Otherwise, the page blocks the operation and shows feedback to the learner. We then translate $M$ into the page. Each operation becomes a direct-manipulation gesture, such as dragging, pouring, or lighting, and each state becomes a visual appearance. Learners therefore perform the experiment themselves instead of replaying a fixed animation.

The generator also receives three design documents $\mathcal{C}$, which the verifier later checks. Unlike the gold specification of the benchmark, these documents are produced by EduVideo2Code itself. The science document attaches video evidence to each state, operation, and precondition of $M$. The layout document gives the object positions and spatial relations. The asset list records the source and license of each apparatus asset, so that every asset in the page can be traced and legally reused. Each asset is taken from a built-in library of code-drawn apparatus, retrieved from external sources, or generated when no suitable asset exists.

\subsection{Execution-Based Verification}
A verifier, separate from the generator, tests each candidate page in a browser. It receives the procedure $P$, the state machine $M$, and the design documents $\mathcal{C}$, and it performs three checks. First, it compares the apparatus and their spatial relations with the video. Second, it attempts invalid operations and checks that the page blocks them with feedback. Third, it replays the demonstrated procedure. After each step, it compares a screenshot and a state summary of the page with the corresponding video evidence. The state summary is a textual record of the current object states that the generated page exposes. The verifier uses the same backbone model as the generator but operates in an independent session without access to the generation history, and it judges the page only from its runtime behavior and the video evidence.

At round $t$, the verifier returns a score $s_t \in [0,1]$ and diagnostic feedback $\mathcal{F}_t$. The page is accepted if $s_t \geq \theta$. Otherwise, the feedback guides a targeted repair: apparatus errors trigger a new asset search, while operation or state-transition errors trigger a re-inspection of the procedure or a revision of the state machine. Verification and repair repeat until the page is accepted or the repair budget of $T_{\max}$ rounds is used up.

\section{Experiments}
\label{sec:experiments}

We compare EduVideo2Code with six baseline models on Lab2Sim. We describe the setup (Section~\ref{sec:experimental-setup}), report the main results (Section~\ref{sec:main-results}), and analyze each component (Section~\ref{sec:ablation}).

\subsection{Experimental Setup}
\label{sec:experimental-setup}

\paragraph{Baselines.} We use six multimodal models as baselines: Seed-2.1-Pro \cite{seed21}, GLM-5.3-Flash \cite{glm53}, GPT-5.6-Sol \cite{gpt56}, GPT-6-Astra \cite{gpt6astra}, DeepSeek-V4.1-Flash \cite{deepseekv41flash}, and Kimi-K3 \cite{kimik3}. Each baseline is prompted to generate the code of an interactive virtual experiment directly from the video. To test EduVideo2Code with different backbones, we instantiate it with GLM-5.3-Flash and with GPT-5.6-Sol.

\paragraph{Implementation Details.} We set the acceptance threshold to $\theta = 0.8$ and the repair budget to $T_{\max} = 10$. Averaged over both backbones, a page undergoes 5.3 repair rounds before it is accepted or the budget is used up. All baselines receive 32 frames sampled uniformly from each video, with the generation instruction in Appendix~\ref{app:prompts}.

\paragraph{Benchmark and Protocol.} We evaluate all systems on all 218 instances of Lab2Sim (Section~\ref{sec:benchmark}). Each model receives the video through its provider-recommended input pipeline, and each generated page runs in a sandboxed Chromium browser at a resolution of $2560\times1440$. To decouple operation from judgment, evaluation is organized into two stages. First, an evaluation agent (\ie Qwen-3.8-Max \cite{qwen38max}) inspects the source code to locate the controls of each annotated step, executes the steps through browser actions such as clicking, hovering, and scrolling, and records the entire interaction at 30 FPS. Second, a separate judge (\ie Gemini-3.8-Flash \cite{gemini38flash}) scores each rubric item by comparing this recording with the video evidence attached to the item. Neither model is among the evaluated systems, and both are applied to all systems under identical settings. State items combine script assertions with visual checks, while constraint items attempt a step before its precondition holds and check that the page blocks it with feedback. On 50 sampled cases, the judge agrees with blind human judgments at a Cohen's $\kappa$ of 0.79. Appendix~\ref{app:evaluation-protocol} gives further details.

\paragraph{Metrics.} We report the four rubric dimensions defined in Section~\ref{sec:benchmark-dimensions}: AF, PE, PCC, and STF. We also report two diagnostic metrics. \emph{Runnable Rate (RR)} is the fraction of pages that load and respond to interaction. \emph{Operation Failure Rate (OFR)} is the fraction of annotated steps that the evaluation agent fails to execute within three attempts (one initial attempt and up to two model-guided retries). A step counts as failed if its target apparatus or control is missing, if the page shows no visible change after the action, or if the required operation is not implemented. OFR is therefore a lenient check: any visible response counts as success. PE is stricter, because it also requires the correct object and the specified result. A page can thus have a low OFR and a low PE at the same time. Higher is better for all metrics except OFR.

\begin{table*}[t]
\centering
\caption{Results on all 218 instances of Lab2Sim (\%). AF, PE, PCC, and STF are the four rubric dimensions; OFR and RR are diagnostic metrics. Subscripts give the change relative to direct generation with the same backbone; for OFR, a negative change is an improvement. The best result in each column is in bold.}
\label{tab:main-results}
\small
\setlength{\tabcolsep}{6pt}
\begin{tabular*}{\textwidth}{@{\extracolsep{\fill}}lcccccc}
\toprule
& \multicolumn{4}{c}{Rubric dimensions} & \multicolumn{2}{c}{Diagnostics} \\
\cmidrule(lr){2-5}\cmidrule(lr){6-7}
Model & AF $\uparrow$ & PE $\uparrow$ & PCC $\uparrow$ & STF $\uparrow$ & OFR $\downarrow$ & RR $\uparrow$ \\
\midrule
\multicolumn{7}{l}{\textit{Direct generation}} \\
Seed-2.1-Pro & 20.62 & 20.11 & 25.14 & 24.65 & 31.09 & 93.12 \\
GLM-5.3-Flash & 40.44 & 22.58 & 16.33 & 34.48 & 12.90 & 94.95 \\
GPT-6-Astra & 41.81 & 24.06 & \textbf{44.83} & 37.88 & 22.28 & 96.79 \\
GPT-5.6-Sol & 36.61 & 32.79 & 32.56 & 41.68 & 28.52 & 96.79 \\
DeepSeek-V4.1-Flash & 59.09 & 36.48 & 36.67 & 41.83 & \textbf{3.14} & \textbf{100.00} \\
Kimi-K3 & 54.23 & 28.57 & 22.22 & 44.86 & 8.04 & \textbf{100.00} \\
\midrule
\multicolumn{7}{l}{\textit{EduVideo2Code}} \\
\quad + GLM-5.3-Flash & \textbf{61.32}\textsubscript{+20.88} & 37.17\textsubscript{+14.59} & 38.33\textsubscript{+22.00} & \textbf{49.63}\textsubscript{+15.15} & 7.10\textsubscript{$-$5.80} & \textbf{100.00}\textsubscript{+5.05} \\
\quad + GPT-5.6-Sol & 60.76\textsubscript{+24.15} & \textbf{41.40}\textsubscript{+8.61} & 33.75\textsubscript{+1.19} & 45.45\textsubscript{+3.77} & 12.38\textsubscript{$-$16.14} & \textbf{100.00}\textsubscript{+3.21} \\
\bottomrule
\end{tabular*}
\end{table*}

\subsection{Main Results}
\label{sec:main-results}

Table~\ref{tab:main-results} reports the results of the six baselines and the two EduVideo2Code variants. We highlight three findings.

\noindent \textbf{Finding 1:} \textit{A high Runnable Rate says little about fidelity.}
All eight systems reach a Runnable Rate above 93\%, yet their rubric scores differ widely. AF ranges from 20.62 to 61.32, and PE ranges from 20.11 to 41.40. Whether a page loads and responds to clicks therefore does not indicate whether it reproduces the experiment. No single system is best on every metric either: the EduVideo2Code variants rank first and second in AF, the GLM-5.3-Flash variant leads in STF, the GPT-5.6-Sol variant leads in PE, and GPT-6-Astra leads in PCC. These results motivate scoring the four dimensions separately.

\noindent \textbf{Finding 2:} \textit{Representing the apparatus is easier than implementing the procedure.}
For every system, AF is higher than PE, although the gap is only 0.51 points for Seed-2.1-Pro. The two dimensions use different items and checks, so their levels are not strictly comparable. Still, the consistent gap suggests that drawing the required apparatus is easier than making every step work. Executing steps is also different from enforcing preconditions. Kimi-K3 has the second-lowest OFR among the baselines (8.04\%) but a PCC of only 22.22\%, while GPT-6-Astra has the highest PCC (44.83\%) despite an OFR of 22.28\%.

\noindent \textbf{Finding 3:} \textit{EduVideo2Code improves both backbones, with gains that vary by dimension.}
Compared with direct generation from the same backbone, both EduVideo2Code variants improve AF, PE, PCC, and STF, reduce OFR, and reach a Runnable Rate of 100\%. AF rises by 20.88 points for GLM-5.3-Flash and by 24.15 points for GPT-5.6-Sol. The PCC gains, however, differ sharply: 22.00 points for GLM-5.3-Flash but only 1.19 points for GPT-5.6-Sol. With GPT-5.6-Sol, EduVideo2Code still trails two direct baselines in PCC (DeepSeek-V4.1-Flash at 36.67 and GPT-6-Astra at 44.83), and its STF (45.45) is close to that of Kimi-K3 (44.86). The gains in PCC are thus the least consistent, and the benefit of the framework depends on both the dimension and the backbone.

\begin{table}[t]
\centering
\caption{Ablation results (\%) on all 218 instances with GLM-5.3-Flash as the backbone. HPE: hierarchical procedure extraction; SMG: state-machine-driven generation; EBV: execution-based verification. Subscripts give the change relative to the full EduVideo2Code; for OFR, a positive change is a degradation.}
\label{tab:ablation}
\small
\resizebox{1.0 \columnwidth}{!}{
\begin{tabular}{lccccc}
\toprule
Variant & AF $\uparrow$ & PE $\uparrow$ & PCC $\uparrow$ & STF $\uparrow$ & OFR $\downarrow$ \\
\midrule
Direct generation & 40.44 & 22.58 & 16.33 & 34.48 & 12.90 \\
\midrule
EduVideo2Code & \textbf{61.32} & \textbf{37.17} & \textbf{38.33} & \textbf{49.63} & \textbf{7.10} \\
\quad w/o HPE & 48.24\textsubscript{$-$13.08} & 23.53\textsubscript{$-$13.64} & 22.03\textsubscript{$-$16.30} & 38.10\textsubscript{$-$11.53} & 11.61\textsubscript{+4.51} \\
\quad w/o SMG & 60.03\textsubscript{$-$1.29} & 30.52\textsubscript{$-$6.65} & 32.20\textsubscript{$-$6.13} & 37.50\textsubscript{$-$12.13} & 10.32\textsubscript{+3.22} \\
\quad w/o EBV & 59.77\textsubscript{$-$1.55} & 25.81\textsubscript{$-$11.36} & 22.41\textsubscript{$-$15.92} & 40.05\textsubscript{$-$9.58} & 14.84\textsubscript{+7.74} \\
\bottomrule
\end{tabular}}
\end{table}

\begin{figure*}[t]
\centering
\includegraphics[width=\textwidth]{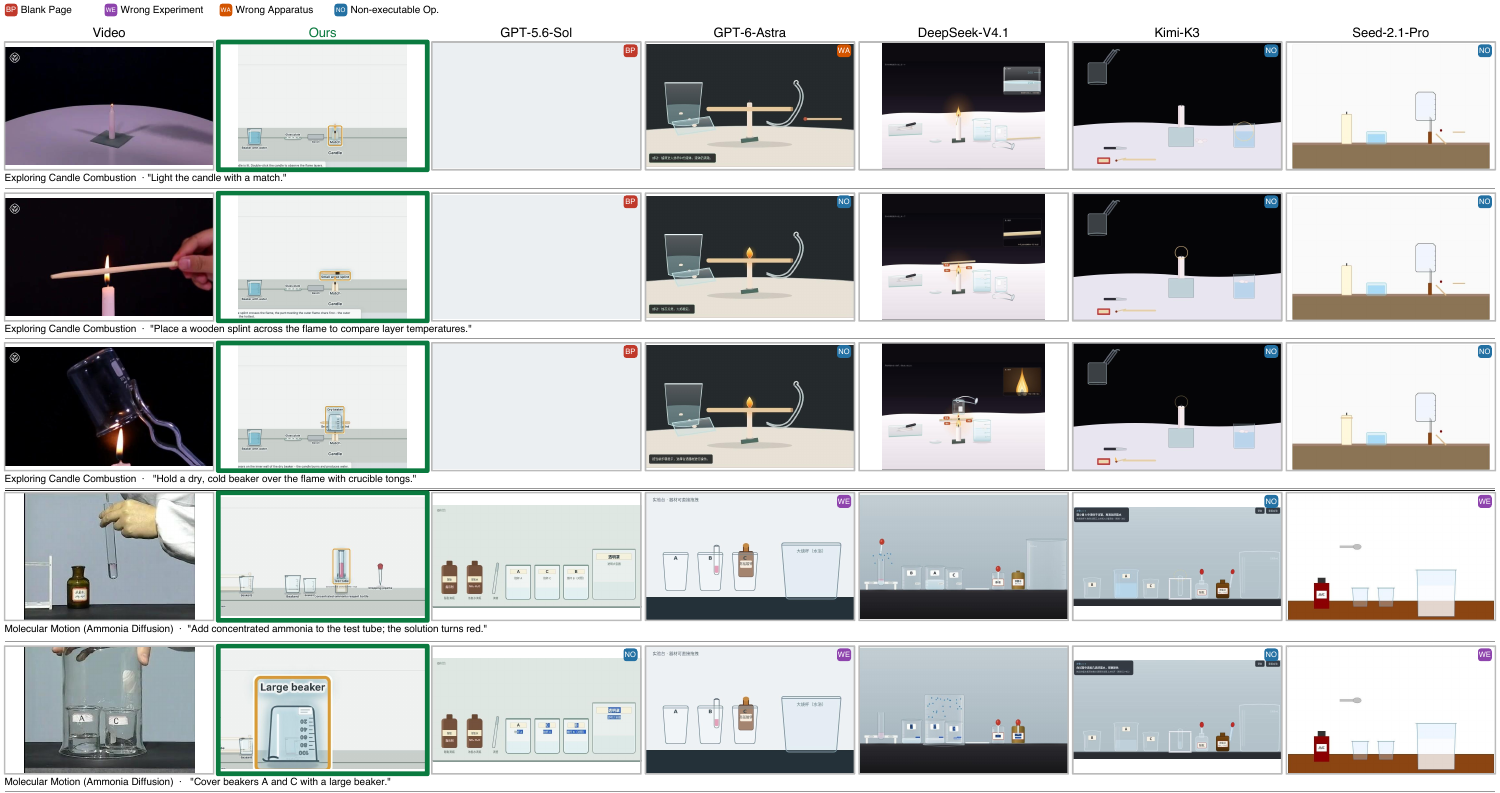}
\caption{Qualitative comparison on two Lab2Sim experiments. Each row shows a key step of the reference video and the state reached in generated environment after the demonstrated operations are replayed. Stickers mark failure types: blank page (BP), wrong experiment (WE), wrong apparatus (WA), and non-executable operation (NO).}
\label{fig:case-study}
\end{figure*}

\subsection{Ablation Study}
\label{sec:ablation}

We ablate the three stages with GLM-5.3-Flash as the backbone. 
As Table~\ref{tab:ablation} shows, removing any stage worsens every metric. Removing HPE causes the largest drops in AF (13.08), PE (13.64 ), and PCC (16.30), and its PE of 23.53 is close to that of direct generation (22.58). Removing SMG causes the largest drop in STF (12.13), only moderate drops in PE and PCC, and almost no change in AF. The state machine thus mainly helps the page produce the correct state transitions. This result only partly matches our motivation for SMG. Explicit preconditions alone do not make the page enforce them: removing SMG lowers PCC by 6.13 points, whereas removing HPE or EBV lowers it by about 16 points. Effective enforcement appears to need preconditions recovered from the video by HPE and checked in the browser by EBV. Removing EBV yields the highest OFR (14.84), which is even higher than that of direct generation (12.90). One likely reason is that the state machine specifies more interactions than a directly generated page, and without execution feedback the broken ones are never repaired. For PE, PCC, and STF, the drops from removing single stages sum to more than the total gain over direct generation, so the stages depend on one another rather than contributing independently.

\subsection{Case Study}
\label{sec:case-study}

Figure~\ref{fig:case-study} presents a qualitative comparison of environments generated for two Lab2Sim experiments, candle combustion and molecular motion, evaluated by replaying the demonstrated operations. We identify four representative failure modes.
\textbf{(i) Blank Rendering.} GPT-5.6-Sol produces a candle-combustion page that loads without runtime errors but displays only ``[object Object]'' instead of the experimental scene. This case illustrates that technical executability does not guarantee meaningful visual reconstruction.
\textbf{(ii) Experiment Substitution.} For molecular motion, GPT-6-Astra and Seed-2.1-Pro generate potassium permanganate and fuchsine diffusion experiments, respectively, rather than reproducing the ammonia and phenolphthalein setup shown in the video. Although both environments are internally coherent, they reflect prior scientific knowledge rather than the specific experimental evidence.
\textbf{(iii) Apparatus Mismatch.} For candle combustion, GPT-6-Astra generates a balance-like stand and a bent tube instead of the demonstrated candle, knife, and crucible tongs. The resulting discrepancy between the generated apparatus and the reference experiment undermines the validity of subsequent interactions.
\textbf{(iv) Non-Executable Operations.} Kimi-K3 and Seed-2.1-Pro generate visually plausible candle scenes, but actions such as lighting the candle, inserting the wooden splint, and positioning the beaker over the flame do not trigger the expected state changes. Similarly, GPT-5.6-Sol renders the molecular-motion apparatus but does not support dragging either beaker to perform the covering step. These cases expose a gap between visual plausibility and functional interaction.

Together, these failure modes reveal distinct challenges in reconstructing interactive experiments, spanning rendering, experiment-level fidelity, apparatus fidelity, and procedural execution. In contrast, EduVideo2Code reproduces the demonstrated operations and their resulting states, including the solution turning red after ammonia is added, by encoding the procedure in an explicit state machine and validating interactions through browser replay. 
However, accurately rendering subtle physical phenomena remains challenging: for example, condensation on the beaker wall is conveyed through text rather than a corresponding visual change.

\section{Conclusion}
\label{sec:conclusion}

We introduced Lab2Sim, a benchmark for reconstructing interactive virtual experiments from real-world experiment videos, and EduVideo2Code, a compile-verify-repair framework for this task. Video-grounded specifications provide concrete targets, while browser execution connects these targets to the behavior of generated environments. Results across six models reveal a persistent gap between producing runnable pages and faithfully reproducing the demonstrated experiment, which EduVideo2Code narrows through explicit state machines and execution-based verification. 
Our work advances the study of translating physical demonstrations into faithful, executable environments.

\section*{Limitations}

Lab2Sim scores a page against what the source video shows, so scientifically expected behavior that the video does not demonstrate is not evaluated. Part of the scoring relies on a model judge, whose agreement with humans has been measured on 50 sampled cases rather than on the full benchmark. The benchmark covers 218 videos in three disciplines from two public platforms, and it does not represent all languages, regions, or laboratory settings. EduVideo2Code requires more model calls than direct generation because of its verification and repair rounds. Finally, we assess the generated pages only with automatic metrics and have not studied their effect on learners, so they should be reviewed by educators before classroom use.

\bibliography{custom}

\appendix

\section{Gold Annotation Guidelines}
\label{app:annotation-guidelines}

\paragraph{Evidence collection.} Annotation proceeds from coarse to fine. The first pass samples 12 frames uniformly over the video and adds frames at scene changes, which are detected at 2 FPS with a scene-change threshold of 0.28, up to 48 frames per video. Candidate frames less than 0.55\,s apart are merged. A frame whose 64-bit average hash differs from that of an earlier frame by at most 6 bits is marked as a near-duplicate and excluded from the review material. The remaining frames are arranged into a contact sheet and into $2\times2$ review sheets, and each tile is labeled with its frame index and timestamp. From these sheets, the assisting model drafts a brief that records the experiment goal, its phases with approximate time ranges, and a checklist of the expected apparatus, each marked as seen, unclear, or missing. Intervals that contain unclear or missing apparatus are re-sampled at a higher rate, and the brief is updated after each round. This loop ends only when every expected apparatus is either clearly visible in a cited frame or confirmed absent after a targeted search of the relevant intervals. A frame that was not sampled is never treated as evidence that an object is absent. An apparatus that remains invisible after the search is labeled \texttt{apparatus\_inferred}, and the searched intervals are recorded with it. For videos edited from still shots, a real state change, such as an object being added to the scene, may leave the average hash unchanged. In such cases, the duplicate filter is bypassed after the frames are confirmed to differ.

\paragraph{Global and local passes.} Based on the brief, a global pass fixes the experiment goal, its phases, and the approximate step boundaries. A local pass then refines each segment of the video into steps. For every step, it records a verb-phrase title, a one-sentence description of the manipulation, quantity, or observation, the start and end times, the apparatus involved, the post-state transitions with their evidence labels, and the cited frames. Full-resolution frames are opened whenever a sheet tile is insufficient to read a color, a liquid level, or a small object. The local results are then merged in temporal order. Table~\ref{tab:gold-example} shows an abridged gold specification and the rubric items derived from it.

\paragraph{Step validity.} A step is either a manipulation that the demonstrator performs on the apparatus or an observation interval in which a phenomenon develops. An observation step must record the observed change. Video packaging such as title cards, section transitions, subtitle frames, and closing screens is excluded. The initial placement of the apparatus is recorded as the initial condition rather than as a step, but assembly actions that actually manipulate the apparatus remain steps. A validator rejects pseudo-steps and non-monotonic timestamps.

\paragraph{Temporal ordering.} Each step starts at the last frame before the action begins and ends at the first frame in which the result is fully visible. Timestamps must be monotonic across the procedure and lie within the video duration. When the video is annotated in segments, a step that spans a segment boundary belongs to the earlier segment.

\paragraph{Entity scope.} Entities record apparatus, reagents, and devices only. The demonstrator, their hands, and the camera viewpoint are not entities. The actor of an operation is the apparatus or tool being manipulated, such as a dropper or a reagent bottle.

\paragraph{State records.} Each state transition is an update to one attribute of a named object. It is recorded as a path from the object to the changed attribute (\eg \texttt{flask.solution.color}), a comparison operator, and a value. The operator is one of \texttt{eq}, \texttt{neq}, \texttt{gt}, \texttt{gte}, \texttt{lt}, \texttt{lte}, and \texttt{between}. Values are semantic tokens, such as \emph{colorless}, \emph{white turbid}, and \emph{lit}, rather than raw measurements. States are recorded per object rather than as a description of the whole scene. Initial conditions use the same format. They are recorded separately from the steps, together with the earliest time at which the apparatus is in place after any title card. Quantities that the footage cannot support, such as an unlabeled molar concentration, are recorded as assumptions with the \texttt{domain} label rather than as state transitions.

\paragraph{Preconditions.} Each precondition is recorded as an executable violation specification. It consists of the operation to attempt and the state overrides under which the precondition does not hold, for example setting the volume of the source container to zero before a pouring step. Each precondition also carries an evidence label and cites the frames in which the required state is visible.

\paragraph{Evidence labels.} Every entity, step, state transition, and precondition carries exactly one of three labels. \texttt{video\_visible} marks a fact that can be seen directly in a cited frame. A fact that is both visible and scientifically predictable is labeled \texttt{video\_visible}. \texttt{apparatus\_inferred} marks a property read from the apparatus context, such as a concentration inferred from a reagent label. It is used only after a targeted search confirms that the value cannot be read directly from the footage. \texttt{domain} marks textbook scientific knowledge that the video does not show. Such claims are flagged as domain-derived and are admitted only after the human review described in Appendix~\ref{app:quality-control}.

\paragraph{Segment evidence.} For each step, 16 frames are sampled uniformly from its annotated time span, including both endpoints, and stitched into a $4\times4$ mosaic in which every tile is labeled with its index and timestamp. The first tile shows the scene just before the operation, and the last tile shows the result. The initial condition receives a mosaic of the still interval between the time at which the apparatus is in place and the start of the first step. These mosaics serve as the video evidence of execution, constraint, and state items.

\begin{table*}[t]
\centering
\caption{Gold specification and derived rubric items for an experiment in which sodium carbonate solution is poured into lime water (abridged). Each row is one gold field. The right columns give the item instantiated from it, how the item is checked, and the evidence it carries. Text is translated from Chinese.}
\label{tab:gold-example}
\small
\renewcommand{\arraystretch}{1.25}
\begin{tabular}{>{\raggedright\arraybackslash}p{0.11\textwidth}>{\raggedright\arraybackslash}p{0.27\textwidth}>{\raggedright\arraybackslash}p{0.27\textwidth}>{\raggedright\arraybackslash}p{0.13\textwidth}>{\raggedright\arraybackslash}p{0.12\textwidth}}
\toprule
Gold field & Annotated content & Derived item (dimension) & Check & Evidence \\
\midrule
Apparatus & \texttt{test-tube-1}, a test tube containing lime water. Visible in frames 4 and 11, boxed on frame 11. & The page presents a test tube containing lime water. (AF) & Judge & Crop and boxed frame \\
Initial condition & \texttt{test-tube-1.color} \texttt{eq} \emph{colorless}, \texttt{video\_visible}. & Before any step, the solution in the test tube is colorless. (STF) & Script and probe & Initial mosaic \\
Step & \texttt{step-02}: pour sodium carbonate solution into the lime water, from 41.5\,s to 47.8\,s. & Step 2 can be performed on the page and takes effect. (PE) & Replay and judge & Mosaic of step 2 \\
State transition & After \texttt{step-02}: \texttt{test-tube-1.color} \texttt{eq} \emph{white turbid}, \texttt{video\_visible}. & After step 2, the solution in the test tube turns white and turbid. (STF) & Script and probe & Mosaic of step 2 \\
Precondition & Pouring requires a non-empty source. Violation: set \texttt{bottle-na2co3.volume\_ml} to 0, then pour. & Pouring from an empty bottle is blocked with feedback. (PCC) & Injected violation and judge & Mosaic of step 2 \\
\midrule
Violation probe & Generated from the step order. & Attempting step 2 without step 1 is blocked. (PCC, \texttt{domain}) & Injected violation and judge & Mosaics of steps 1 and 2 \\
\bottomrule
\end{tabular}
\end{table*}

\section{Rubric Instantiation Templates}
\label{app:instantiation-templates}

Candidate rubric items are produced by the fixed field-to-item templates in Table~\ref{tab:instantiation-templates}. Each field in the gold specification yields exactly one item of its own type. Different fields can still describe the same requirement, for example a precondition and the post-state of the preceding step, and such duplicates are removed during human review. Question text is composed only from gold fields. Entity labels, contents, step numbers, titles, and descriptions are copied verbatim, so instantiation adds no content beyond the gold specification.

\paragraph{Item format.} Each item records its identifier, dimension, question, gold answer, check specification, evidence, a pointer to its source field, and the evidence label inherited from that field. The check specification is one of three types: a model judgment, a script check, or a dual-track check that combines a script assertion with a visual probe question. Table~\ref{tab:gold-example} gives one item of each type.

\paragraph{Evidence mounting.} Evidence is attached by the same templates. An identity item carries a close-up crop of the apparatus and the full frame with the apparatus boxed, both cut from the bounding box in the gold specification. An execution item carries the segment mosaic of its step. A constraint item carries the mosaic of the step whose precondition it violates. A state item carries the mosaic of the step that produces the state, whose last tile shows the result, and an initial-condition item carries the initial mosaic. When a mosaic is unavailable, the item falls back to the frames cited by its source field.

\paragraph{Violation probes.} The only items not derived one-to-one from the gold fields are the violation probes. A skip probe is generated for every step after the first, and it attempts the step without first performing the preceding one. A reorder probe is generated for every pair of adjacent steps whose operations involve different actors, and it performs the two steps in swapped order. An overdose probe is generated for every step with a numeric quantity, and it multiplies the quantity by five. A substitute probe is generated for every step whose target is a container with another container of the same kind in the scene, and it applies the step to that other container. A probe passes when the page blocks the violating operation with textual feedback. It fails when the page executes the operation silently or crashes. Probes are labeled \texttt{domain} and count toward PCC together with the constraint items.

\begin{table*}[t]
\centering
\caption{Field-to-item templates that map gold fields to candidate rubric items. Dim.\ gives the rubric dimension of each item. Violation probes are labeled \texttt{domain} and count toward PCC.}
\label{tab:instantiation-templates}
\small
\renewcommand{\arraystretch}{1.2}
\begin{tabular}{>{\raggedright\arraybackslash}p{0.14\textwidth}>{\raggedright\arraybackslash}p{0.29\textwidth}>{\raggedright\arraybackslash}p{0.25\textwidth}>{\raggedright\arraybackslash}p{0.18\textwidth}l}
\toprule
Gold field & Item & Check & Evidence & Dim. \\
\midrule
Apparatus entity & The named apparatus and its contents are present and correctly identified & Judge decision on attached crops & Boxed crop and scene image & AF \\
Procedure step & The step can be performed on the page and takes effect & Replay of the step & Segment mosaic of the step & PE \\
Precondition & Attempting the step without its precondition is blocked with feedback & Injected violation, expects blocking with feedback & Mosaic of the involved step & PCC \\
Step state transition & After the step, the object attribute has the required value & Script assertion with a visual check & Segment mosaic ending at the result & STF \\
Initial condition & Before any step, the object attribute has the required value & Script assertion with a visual check & Mosaic of the initial still interval & STF \\
Violation probe & A skipped, reordered, overdosed, or misapplied step is blocked & Injected violation, expects blocking or an explicit invalid-result warning & Mosaics of the involved steps & PCC \\
\bottomrule
\end{tabular}
\end{table*}

\paragraph{Rubric statistics.} The frozen rubric contains 11,881 items in total. Of these, 15.54\% are identity items (AF), 13.03\% are execution items (PE), 42.57\% are state items (STF), and 28.86\% count toward PCC, of which 8.10\% are constraint items and 20.76\% are violation probes. Each step yields one execution item and each state transition yields one state item, so the PE and STF shares correspond to the 7.1 steps and 23.2 state transitions per instance in Table~\ref{tab:benchmark-statistics}. The main scores include items based on \texttt{video\_visible} elements and the violation probes. The other items labeled \texttt{apparatus\_inferred} or \texttt{domain} are released with the rubric but excluded from the main scores.

\section{Quality Control}
\label{app:quality-control}

\paragraph{Gold validation.} Each gold specification must pass a schema validator before any evidence is derived from it. The validator rejects references to frames that are not in the evidence manifest (the list of sampled frames that the specification cites), entity identifiers that are duplicated or undeclared, bounding boxes that are not normalized or extend beyond the frame, comparison operators and evidence labels outside the allowed sets, steps without a title or cited frames, step timestamps that are not monotonic or exceed the video duration, and pseudo-steps such as title cards or pure equipment display. It also requires every specification to record the assisting model and the human verifier.

\paragraph{Gold review.} Human reviewers verify every drafted specification against the source video before rubric items are instantiated from it, which takes about 13 minutes per video on average. The review covers every element of the draft. For each apparatus entity, step, state transition, and precondition, the reviewer checks the element against its cited frames, confirms its evidence label, and removes anything the video does not support. The same five reviewers who later perform item filtering carry out this review. Each specification is checked by two reviewers, and a third reviewer decides when the two disagree. About 20\% of the drafted elements are corrected or removed in this pass.

\paragraph{Apparatus localization.} Every \texttt{video\_visible} entity has a bounding box on one of its cited frames. Localization and verification are two separate vision calls. The first call regresses a tight box around the named apparatus on the clean frame. The second call receives the frame with the box drawn on it and identifies, without access to the first answer, which object lies inside the box. A box fails verification when it is empty, encloses the wrong object, truncates the apparatus, or lies on a frame in which the apparatus is not visible. A failed box is re-predicted with the verifier's feedback for at most three rounds. After that, the entity loses its localization evidence and is downgraded to \texttt{apparatus\_inferred}, that is, it is treated as inferred from context rather than directly observed. No unverified box therefore enters the gold specification.

\paragraph{Consensus sampling.} The assisting model runs the step-level annotation pass three times for every instance. For most fields, the majority reading is kept, and the agreement for each field is recorded in the annotation metadata. The exception is a directly observable reading, such as a scale value. If the runs disagree on such a reading, the full-resolution frame is inspected instead of taking a vote.

\paragraph{Item filtering.} Human reviewers work on a page that shows each candidate item next to its visual evidence. Items are kept by default. Each removal must record one reason: ambiguity, unanswerable from the attached evidence, insufficient support in the gold specification, duplication of another item, or another issue. Five reviewers take part: one doctoral student, two master's students, and two undergraduate students. Each item is checked by two reviewers. When the two disagree, a third reviewer decides, so the final decision follows the majority of three. About five percent of the candidate items are removed in this pass.

\paragraph{State checks and tolerances.} Because required states are semantic tokens, most state items are checked by qualitative equality and need no numeric tolerance. When a comparison is quantitative, the assertion uses an ordinal comparison rather than exact numeric equality. Each state item is checked as a post-condition at the step that produces the state.

\paragraph{Judge calibration.} To validate the automatic judge, we sample 50 video cases uniformly at random. The same five reviewers answer the 2983 kept items of these cases blind, without access to the judge verdicts. We compute the agreement between the judge and the human answers over these items, which reaches a Cohen's $\kappa$ of 0.79. Scaling this calibration to the full benchmark remains future work.

\paragraph{Reviewer instructions.} Before the review begins, each reviewer receives written instructions and a walkthrough of the review page. The instructions consist of the annotation guidelines in Appendix~\ref{app:annotation-guidelines}, the review rules for gold specifications described above, and the list of removal reasons used in item filtering. The review page shows each candidate item next to its visual evidence and records one removal reason for every removed item. The task consists only of watching public instructional experiment videos and judging annotations on a computer. Reviewers are informed before they start that it involves no risk beyond ordinary computer work. No personal information about the reviewers is collected or released.

\paragraph{Reviewer recruitment and compensation.} The five reviewers are students recruited from our institution, namely the doctoral, master's, and undergraduate students described in item filtering. Before participating, each reviewer is told the purpose of the annotation and how the annotations will be released, and gives consent. Participation is voluntary, reviewers may stop at any time without penalty, and participation has no effect on their coursework or academic evaluation. The trained reviewers are compensated at a rate of \$15 per hour, which is above the local minimum wage standard. Given the review time of about 13 minutes per video, we consider this payment adequate for the workload.

\paragraph{Coverage and freezing.} Before release, a coverage report lists every step without a kept execution or state item, every entity without a kept identity item, and every precondition whose item was removed. Approved items are frozen into a versioned rubric that does not change during reporting. Any later correction to the gold specification regenerates the evidence and the rubric under a new version, and the previous version is archived.

\section{Evaluation Protocol Details}
\label{app:evaluation-protocol}

\paragraph{Evaluated instances.} All systems are evaluated on the full set of 218 instances, and no instance is excluded for any system.

\paragraph{Generation inputs.} Each baseline receives 32 frames sampled uniformly over the full duration of each video, together with the generation instruction shared by all instances (Appendix~\ref{app:prompts}). The frames are passed to each model through its provider-recommended image input, and they are the only visual input that the model receives.

\paragraph{Runnable check.} Each page is first loaded without any instrumentation. It counts as runnable if it raises no fatal runtime error and its stage renders content, namely an SVG element, a canvas element, or more than 50 characters of text, within 15\,s of loading. Rubric items are scored only for runnable pages. The items of a non-runnable page are marked unscorable, so the page lowers the Runnable Rate but does not enter the four rubric dimensions.

\paragraph{Operation.} The evaluation agent executes the annotated steps in their annotated order on a single page session, so that each step starts from the state produced by the previous ones. Before each step, the agent inspects the source code of the rendered page to locate the controls of the step. Interactive elements are extracted from the page source as those with a pointer or grab cursor, a click handler, or an ARIA role, and each element is recorded with its tag, its text, and its center coordinates. At every turn, the agent receives a screenshot of the page, its visible progress text, the actions it has already taken, and this element list. The agent returns one action per turn: a click, a drag between two points, a completion signal, or a failure signal. A drag is executed as a sequence of eight intermediate mouse movements, so that pages implementing drag-and-drop receive realistic events. Each step is limited to six turns.

\paragraph{Turn budget and retries.} The six-turn budget bounds a single attempt on a step, whereas the three attempts in the OFR definition count retries. One attempt gives the agent one six-turn budget, and the attempt ends when the agent signals completion or failure or when the budget is used up. A step that is judged as not executed receives up to two model-guided retries, each with a fresh six-turn budget, so a step consumes at most 18 turns. A step counts as an operation failure only when all three attempts fail, which matches the definition of OFR in Section~\ref{sec:experimental-setup}.

\paragraph{Item scoring.} Each item receives a score of 1, 0.5, or 0, or is marked unscorable. Unscorable items are excluded from the denominator of their dimension.
\begin{itemize}
\item \textbf{AF.} The judge receives the initial screenshot of the page together with the close-up crop and the boxed frame of the apparatus. It returns \textit{match}, \textit{mismatch}, or \textit{uncertain}, which yield 1, 0, and unscorable, respectively.
\item \textbf{PE.} The judge compares the screenshots taken before and after the agent performs the step. The item scores 1 if the core operation of the step is accepted and produces a state change, a phenomenon, or a success message. Highlighting or selection alone does not count. A step for which the agent finds no corresponding control scores 0.
\item \textbf{PCC.} The agent attempts the step while its precondition does not hold, for example by skipping the step that establishes it. The judge receives the resulting screenshot and the visible page text and decides whether the page blocked the attempt with feedback or executed it silently. When several page operations are compatible with the annotated step, all of them are attempted, and the item scores 1 if any of them is blocked.
\item \textbf{STF.} State items are scored on two tracks after the steps up to the target step have been replayed. When the page exposes the state of the object, a script assertion reads the attribute and compares it with the gold value using the annotated operator. Independently, the judge answers the probe question from a screenshot taken at the same point. The item scores 1 when both tracks pass, 0.5 when exactly one passes, and 0 when neither passes. Only a score of 1 counts as a pass in aggregation. When the page does not expose the state, or when the gold value is a free-text token with no counterpart in the page's state vocabulary, the item is scored by the probe alone.
\item \textbf{Violation probes.} Probes are scored in the same way as constraint items and pooled with them in PCC.
\end{itemize}

\paragraph{Judge configuration.} The judge is Gemini-3.8-Flash with temperature 0 and at most 1,024 output tokens. Images are passed as base64-encoded data. Every verdict is returned as JSON and cached under a hash of the model name and the full message, so re-running the evaluation reproduces the same verdicts. Failed requests are retried up to five times with exponential backoff.

\paragraph{Aggregation.} Each dimension is aggregated over the pooled items of all evaluated instances. Its score is the fraction of scored items that pass, where unscorable items are excluded from both the numerator and the denominator. Because the judge runs at temperature 0 and its verdicts are cached, every reported number comes from a single evaluation run.

\section{Generation, Agent, and Judge Prompts}
\label{app:prompts}

This section reproduces the prompts used for generation and evaluation. The evaluation prompts are issued in Chinese, and we give their English translations. Placeholders in braces, such as \texttt{\{task\}}, are filled in for each item at run time.

\subsection{Generation Instruction}

All baselines receive the instruction in Figure~\ref{fig:prompt-generation}. It is identical across instances and contains no information about the specific experiment, so the video is the only source of experimental content. The instruction refers to the input as a video file, and in our experiments this input is provided as the 32 uniformly sampled frames described in Appendix~\ref{app:evaluation-protocol}. The instruction was written in Chinese and English with equivalent content, and we reproduce the English version verbatim.

\begin{figure*}[p]
\centering
\begin{taskbox}[title={Generation instruction: task, input, and output}]
You are a front-end engineer with a background in scientific simulation.
Watch the real classroom/lab experiment video below and build an
interactive web simulation that lets a learner perform the demonstrated
experiment hands-on. Deliver a manipulable experiment -- not a video
summary or an animation replay.

[Input]
1. An experiment video file (MP4 / MOV / WebM) -- the single source of
   truth.
2. (Optional) Video metadata or an ASR transcript -- hints only, not
   ground truth.

[Output]
1. A standard front-end npm project (Vite + vanilla JS/TS recommended, or
   your preferred stack): `npm run dev` starts the interactive simulation
   locally, and `npm run build` produces a static deployable output.
   Apparatus is drawn with SVG or Canvas.
2. README.md (short) -- how to run the project, the scientific
   assumptions made, anything inferred because it was unclear in the
   video, and the video timestamp of each experiment step.
\end{taskbox}
\begin{taskbox}[title={Generation instruction: requirements and completion criteria}]
[Requirements]
1. Apparatus fidelity: every apparatus matches what appears in the video
   (name, shape, function). Never substitute a look-alike, and never
   invent apparatus that is not in the video.
2. Direct manipulation: experiment operations are performed by dragging
   or clicking the apparatus itself (pour, stir, open a valve, place).
   No buttons for experiment operations -- buttons are only for reset,
   help, and other non-experiment UI.
3. State & feedback: learner actions genuinely mutate scientific state
   (volume, color, reading, temperature, ...) and reproduce the phenomena
   shown in the video. Include goal guidance, success/failure feedback,
   and one-click reset. At least one key action must be performed by the
   learner and affect the outcome.
4. Faithfulness: assert only what is visible in the video and keep the
   step order identical to it. Put anything uncertain into the README as
   an assumption. Prefer an honest but incomplete simulation over a
   polished but unrelated one.
5. Local runtime: dependencies are installed via npm only; the running
   app must not hotlink any CDN scripts, fonts, or images -- every asset
   is bundled inside the project. Never use video frames as the stage
   background.

[Done means]
- Faithful to the video (apparatus, steps, phenomena);
- Interactive (actions directly change state and affect the outcome);
- Scientifically correct (state-change rules are reasonable);
- `npm run dev` starts cleanly on a local machine, and every experiment
  step can be completed end to end.
\end{taskbox}
\caption{Generation instruction shared by all baselines and all instances.}
\label{fig:prompt-generation}
\end{figure*}

\subsection{Evaluation Agent}

The agent receives the prompt in Figure~\ref{fig:prompt-agent} at every turn, together with the current screenshot. The \texttt{\{elements\}} field is filled with the interactive elements extracted from the page source (Appendix~\ref{app:evaluation-protocol}). It executes one annotated step at a time and returns a single action per turn.

\begin{figure*}[t]
\centering
\begin{taskbox}[title={Evaluation agent: per-turn instruction}]
You are operating an educational virtual-experiment webpage
(viewport {W}x{H} pixels; coordinates are pixels).
Task: {task}
Actions you have already taken (most recent last): {history}
Interactive elements on the page (center coordinates; choose targets
from these first):
{elements}
Current progress text on the page: {progress}
The current screenshot is attached. A step usually takes 1-3 actions.
If the page progress or state already shows the target effect, reply
done immediately.
Reply with exactly one JSON action per turn and nothing else:
{"action": "click", "x": 123, "y": 456}
{"action": "drag", "x1": .., "y1": .., "x2": .., "y2": ..}
{"action": "done", "note": "what was done and how the page reacted"}
{"action": "fail", "note": "why the step cannot be completed"}
Hint: experimental operations usually drag an apparatus above its target
or click a stopper, valve, switch, or the apparatus itself. Do not use
non-experimental buttons (reset, help), and do not repeat an action you
have already taken.
\end{taskbox}
\caption{Per-turn prompt of the evaluation agent.}
\label{fig:prompt-agent}
\end{figure*}

\subsection{Judge Prompts}

The judge receives one of the prompts in Figure~\ref{fig:prompt-judge} for each item, depending on its dimension. Each prompt asks for a JSON verdict, which is parsed into the item score described in Appendix~\ref{app:evaluation-protocol}.

\begin{figure*}[t]
\centering
\begin{minipage}[t]{0.49\textwidth}
\begin{taskbox}[title={AF judge: apparatus identity},equal height group=judgeA]
You are a strict reviewer of educational virtual-experiment webpages.
Decide whether the [generated page] presents the specified apparatus.
Question: {question}
The first image is the initial screenshot of the generated page. The
other images show this apparatus in the real experiment video
(close-up crop and boxed frame).
Answer only in JSON:
{"verdict": "MATCH" | "MISMATCH" | "UNCERTAIN", "reason": "one sentence"}
\end{taskbox}
\end{minipage}\hfill
\begin{minipage}[t]{0.49\textwidth}
\begin{taskbox}[title={PE judge: step execution},equal height group=judgeA]
Compare the [before] and [after] screenshots of the same virtual-
experiment page.
Operation to perform: {task}
Criterion: if the [core action] of the operation is accepted by the page
and takes effect (a corresponding state change, phenomenon, or success
message appears), answer accepted=true. Other details of the description
that are not completed do not affect the decision, since the page may
divide steps at a different granularity than the task description.
Only highlighting, selection, or no change at all counts as false.
Answer only in JSON:
{"accepted": true | false, "reason": "one sentence of evidence"}
\end{taskbox}
\end{minipage}

\begin{minipage}[t]{0.49\textwidth}
\begin{taskbox}[title={PCC judge: precondition enforcement},equal height group=judgeB]
This is a screenshot of a virtual-experiment page after an [invalid
operation] was attempted (the visible page text is attached).
Invalid operation: {task}
The page should [block the operation and indicate the missing condition
or hazard], rather than execute it normally.
Decide the page behavior and answer only in JSON:
{"blocked": true | false, "reason": "one sentence of evidence, citing
the visible message or state"}
\end{taskbox}
\end{minipage}\hfill
\begin{minipage}[t]{0.49\textwidth}
\begin{taskbox}[title={STF judge: state probe},equal height group=judgeB]
This is a screenshot of an educational virtual-experiment page in a
certain state. Probe question: {question}
Candidate answers: yes/no/uncertain.
Observe what the page currently shows and answer only in JSON:
{"answer": "yes" | "no" | "uncertain", "reason": "one sentence"}
\end{taskbox}
\end{minipage}
\caption{Judge prompts for the four rubric dimensions.}
\label{fig:prompt-judge}
\end{figure*}

\section{Additional Case Studies}
\label{app:case-studies}

\begin{figure*}[p]
\centering
\includegraphics[width=\textwidth]{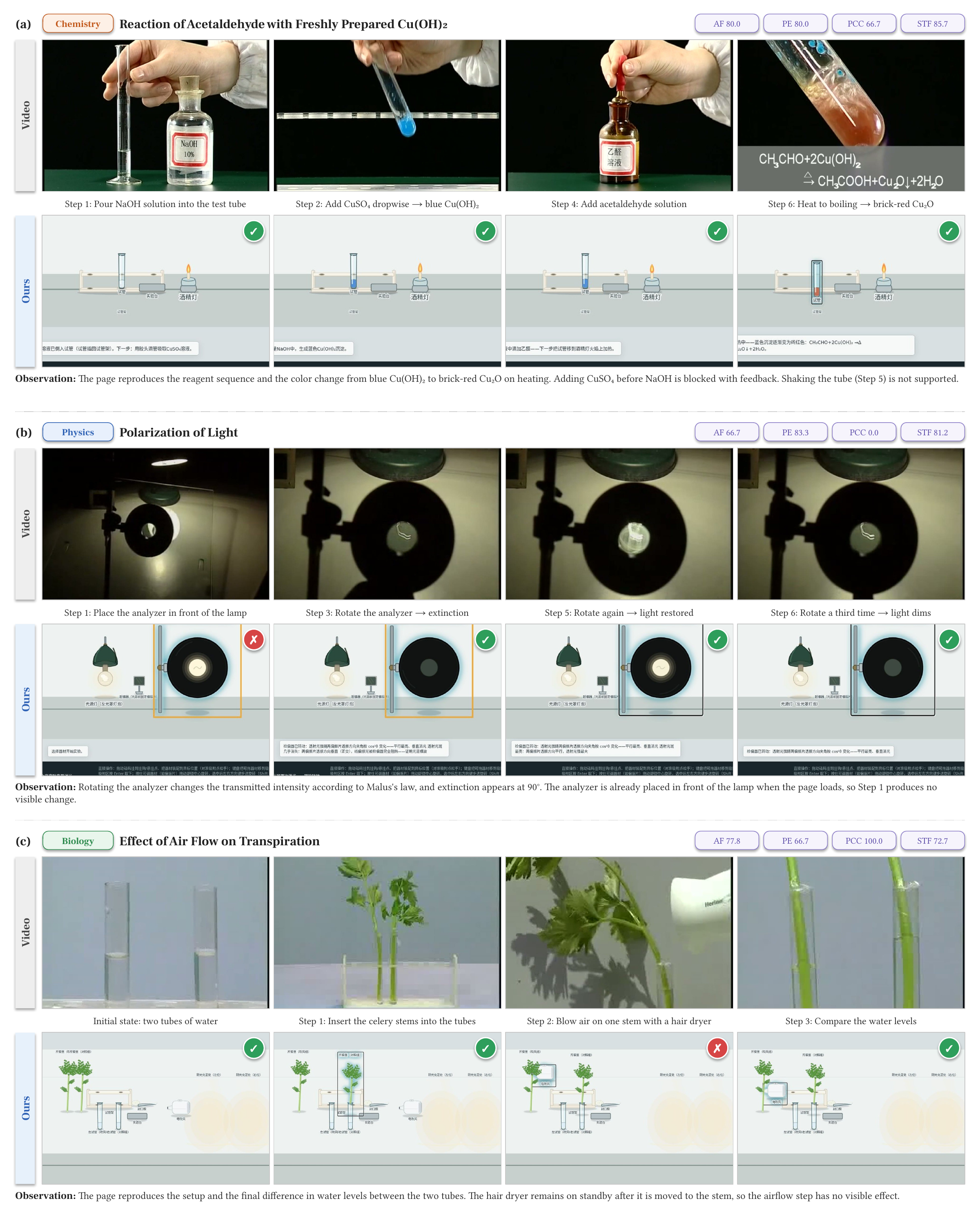}
\caption{Additional case studies of EduVideo2Code with GLM-5.3-Flash (Part I). For each instance, the top row shows key frames of the source video and the bottom row shows the generated page after the evaluation agent performs the same step. \ding{51}/\ding{55} marks the judge verdict on the procedure execution item of each step.}
\label{fig:app-cases-1}
\end{figure*}

\begin{figure*}[p]
\centering
\includegraphics[width=\textwidth]{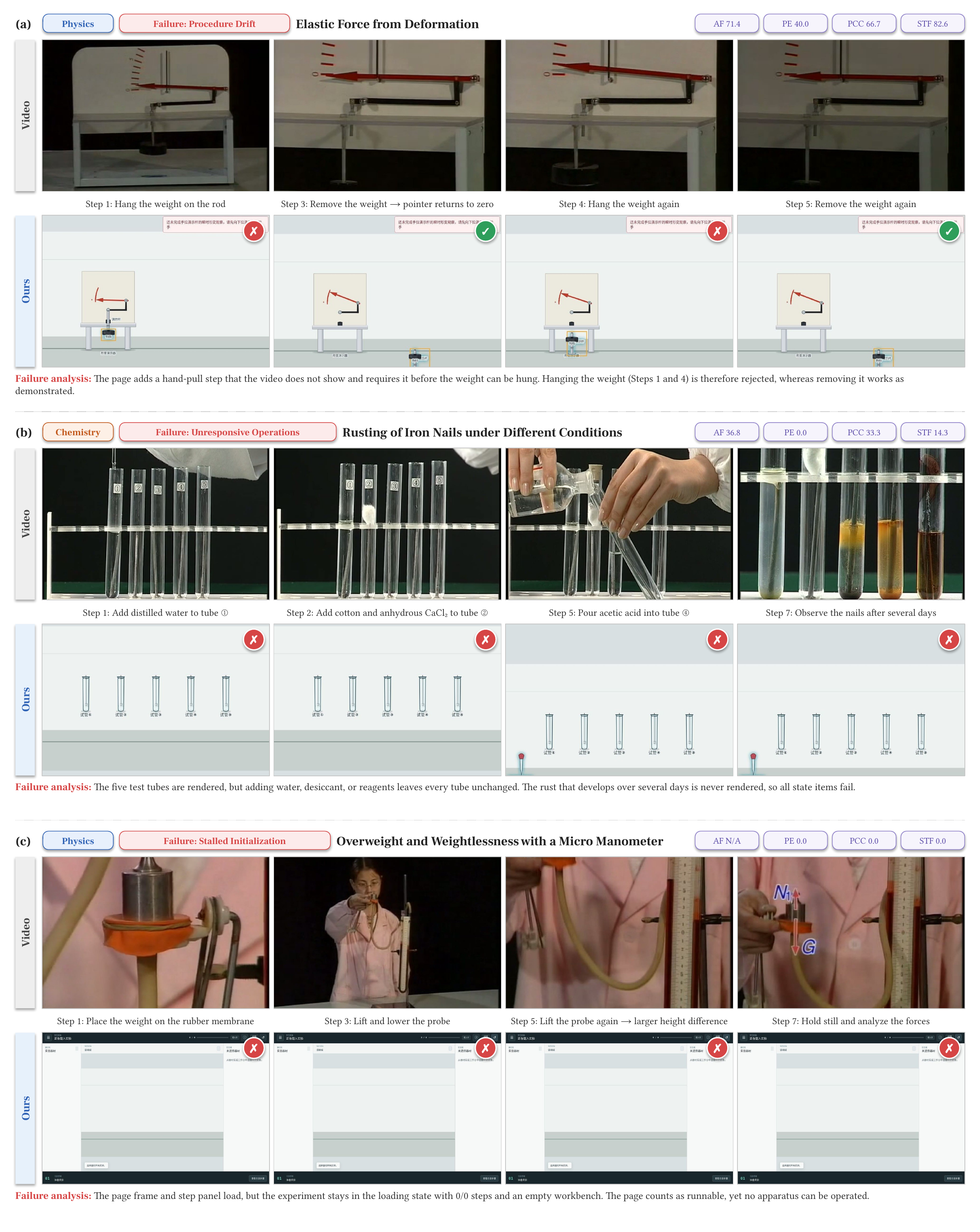}
\caption{Additional case studies of EduVideo2Code with GLM-5.3-Flash (Part II): failure cases. (a) Procedure drift, where the page imposes a step absent from the video. (b) Unresponsive operations, where no operation changes the scene. (c) Stalled initialization, where the experiment never leaves its loading state.}
\label{fig:app-cases-2}
\end{figure*}

Figures~\ref{fig:app-cases-1} and~\ref{fig:app-cases-2} show six further Lab2Sim instances, none of which appears in Section~\ref{sec:case-study}. All pages are generated by EduVideo2Code with GLM-5.3-Flash. For each instance, the top row shows key frames of the source video for four annotated steps, and the bottom row shows the page after the evaluation agent performs the same steps. Each page frame is marked with the judge verdict on the procedure execution item of that step, and the header of each instance gives its four rubric scores.

\paragraph{Successful reconstructions.} Figure~\ref{fig:app-cases-1} covers one instance from each discipline. In the acetaldehyde experiment, the page executes the reagent sequence and turns the solution from blue to brick red on heating, and it blocks the addition of CuSO$_4$ before NaOH with textual feedback. In the polarization experiment, rotating the analyzer changes the transmitted intensity according to Malus's law, and extinction appears at 90$^\circ$. In the transpiration experiment, the page reproduces the setup and the final difference in water levels between the two tubes. Even these pages miss individual steps. The acetaldehyde page offers no interaction for shaking the test tube. The analyzer in the polarization page is already in place when the page loads, so Step~1 produces no visible change. The hair dryer in the transpiration page remains on standby after it is moved to the stem, so the airflow step has no visible effect.

\paragraph{Failure cases.} Figure~\ref{fig:app-cases-2} shows three failure modes that are not covered in Section~\ref{sec:case-study}.
\textbf{(i) Procedure Drift.} In the deformation experiment, the page adds a hand-pull step that the video does not show and makes it a precondition for hanging the weight. Hanging the weight in Steps~1 and~4 is therefore rejected, whereas removing the weight works as demonstrated. The page does enforce a precondition, but the precondition comes from the model's own version of the procedure rather than from the video, so the constraint blocks a demonstrated step.
\textbf{(ii) Unresponsive Operations.} In the rusting experiment, the page renders the five test tubes, but adding water, desiccant, or reagents leaves every tube unchanged, and the rust that develops over several days is never rendered. The page thus reproduces part of the apparatus but none of the state transitions, which yields a PE of 0 and an STF of 14.3.
\textbf{(iii) Stalled Initialization.} In the manometer experiment, the page frame and the step panel load, but the experiment remains in its loading state with zero steps and an empty workbench. The page raises no fatal error and is counted as runnable, yet no apparatus can be operated. Like the blank rendering in Section~\ref{sec:case-study}, this case shows that a runnable page is not necessarily a usable experiment.

\end{document}